\documentclass[aps,prl,twocolumn,floatfix,superscriptaddress,preprintnumbers]{revtex4-2}

\pdfoutput=1
\usepackage{amsmath,amssymb}
\usepackage{amsfonts}
\usepackage{bm,bbm}
\usepackage{graphicx}
\usepackage{mathrsfs}
\usepackage{slashed}
\usepackage{booktabs}
\usepackage{tabu}
\usepackage[dvipsnames]{xcolor}
\usepackage[normalem]{ulem}
\usepackage{tikz}
\usepackage{soul}
\usepackage[colorlinks=true,linkcolor=blue]{hyperref}
\usepackage{xcolor}
\usepackage{ulem}
\usepackage{array}
\usepackage{verbatim}
\usepackage{epsfig}
\usepackage{multirow}

\begin{document}

\preprint{ADP-22-4/T1175}

\title{Sensitivity of Parity-violating Electron Scattering to a Dark Photon}

\author{A. W.~Thomas}
\affiliation{ARC Centre of Excellence for Dark Matter Particle Physics and CSSM, Department of Physics, University of Adelaide, SA 5005, Australia}
\date{\today}

\author{X. G.~Wang}
\affiliation{ARC Centre of Excellence for Dark Matter Particle Physics and CSSM, Department of Physics, University of Adelaide, SA 5005, Australia}

\author{A. G.~Williams}
\affiliation{ARC Centre of Excellence for Dark Matter Particle Physics and CSSM, Department of Physics, University of Adelaide, SA 5005, Australia}

\begin{abstract}
We explore the sensitivity of the parity-violating electron scattering (PVES) asymmetry in both elastic and deep-inelastic scattering to the properties of a dark photon. Given advances in experimental capabilities in recent years, there are interesting regions of parameter space where PVES offers the chance to discover new physics in the near future. There are also cases where the existence of a dark photon could significantly alter our understanding of the structure of atomic nuclei and neutron stars as well as parton distribution functions. 
\end{abstract}

\date{\today}
\maketitle


{\em Introduction:}  Parity-violating electron scattering (PVES) has been proposed as an important new tool for testing the Standard Model (SM), probing new physics and studying hadron and nuclear structure.   

Elastic PVES experiments have been used to measure neutral weak form factors.
The Qweak Collaboration~\cite{Qweak:2018tjf} recently provided an important test of the SM by extracting the proton weak charge from a  high-precision measurement of the parity-violating asymmetry in the scattering of polarized electrons on protons.
The PREX experiments provided precise measurements of the parity violating asymmetry in electron scattering from a $^{208}$Pb target~\cite{Abrahamyan:2012gp, PREX:2021umo}. As the weak charge of the neutron is much larger than that of the proton, this effectively measured the distribution of neutrons, leading to a determination of the neutron skin thickness, the central nuclear density and the density dependence of the symmetry energy. The neutron radius deduced from this measurement was significantly larger than expected from structure calculations~\cite{Reinhard:2021utv,Martinez:2018xep}, although in the present context  it is interesting to note that a recent 
study~\cite{Corona:2021yfd} showed that this tension might be relieved by a small change in the Weinberg angle.

Parity-violating deep-inelastic scattering (PVDIS) has proven particularly valuable in testing the SM. The first PVES experiment in the DIS region was performed at the Stanford Linear Accelerator Center (SLAC) on a deuteron target~\cite{Prescott:1978tm, Prescott:1979dh},
providing important early confirmation of the SM.
A more precise measurement was carried out at the Thomas Jefferson National Accelerator Facility (Jefferson Lab)~\cite{PVDIS:2014cmd, Wang:2014guo},
providing direct evidence of the non-zero $C_{2q}$ couplings predicted by the electroweak theory.
Recently, it was proposed to measure the single-spin asymmetry in PVDIS with $b$-tagged jets at HERA and EIC to probe the $Zb\bar{b}$ anomalous couplings~\cite{Yan:2021htf, Li:2021uww}.
In addition, PVDIS promises to play a vital role in exploring the partonic structure of the nucleon and nuclei. For example, it promises a model independent method to extract the 
fundamental~\cite{Close:1988br} ratio $d/u$ at large Bjorken-$x$~\cite{Hobbs:2008mm}. It also offers new insight into the isovector nuclear EMC effect~\cite{Cloet:2012td,Cocuzza:2021rfn} and confirmation of one proposed correction~\cite{Bentz:2009yy} to the NuTeV 
measurement of the Weinberg angle and the associated anomaly~\cite{NuTeV:2001whx}.

In atomic physics too, PVES has been employed as a probe of new
 physics beyond the SM using atomic parity violation~\cite{Marciano:1990dp}.
Most recently, the potential impact of future PVES experiments on new physics was investigated in the framework of SM effective field theory (SMEFT)~\cite{Boughezal:2021kla, Bischer:2021jqn}.
The search for physics beyond the SM in PVES is also one of the primary goals of the ongoing scientific program of the 12 GeV Continuous Electron Beam Accelerator Facility (CEBAF) at Jefferson Lab~\cite{Arrington:2021alx}, especially the SoLID Collaboration~\cite{Chen:2014psa}.

In this Letter, we propose that PVES offers particularly promising opportunities to search for physics beyond the SM in the form of a dark photon.
Furthermore, we investigate the sensitivity of the PVES asymmetry to the dark photon parameters and the potential implications for the interpretation of experiments already carried out. These investigations reveal potentially sizeable corrections over a wide range of momentum transfer from DIS near the $Z$ mass to much lower energy elastic scattering from Pb. In the latter case the correction could significantly alter the interpretation of the PREX experiment, with implications from nuclear structure to neutron stars.


{\em PVES asymmetry:}  In scattering of longitudinally polarized electrons on an unpolarized target, parity-violation effect is characterized by the asymmetry between left- and right-handed electrons
\begin{equation}
A_{\rm PV} = \frac{\sigma_R - \sigma_L}{\sigma_R + \sigma_L} \, ,
\end{equation}
where $\sigma_{R,L} = d^2 \sigma_{R,L}/d\Omega dE'$ are the double differential cross sections of right-handed (R) and left-handed (L) electrons, respectively. 

For elastic scattering, this asymmetry can be expressed in terms of the weak form factor $F_W$ and the charge form factor $F_C$~\cite{Horowitz:1999fk}, 
\begin{equation}
A^{\rm el}_{\rm PV} = \frac{G_F Q^2 |Q^{(W)}_{N,Z}|}{4\sqrt{2} \pi \alpha Z} \frac{F_W(Q^2)}{F_C(Q^2)} \, ,
\end{equation}
where $G_F = 1.1663787\times 10^{-5} {\rm GeV}^{-2}$ is the Fermi constant, and $Q^{(W)}_{N,Z}$ is the weak charge of the target nucleus with $N$ neutrons and $Z$ protons.

In the case of deep inelastic scattering (DIS), the beam asymmetry has a simple form in leading order of one-photon and one-$Z^0$ exchanges~\cite{Brady:2011uy}
\begin{equation}
A^{\rm DIS}_{\rm PV} = \frac{G_F Q^2}{4\sqrt{2}(1+Q^2/M_Z^2) \pi \alpha} 
\Big[ a_{1} + \frac{1 - (1 - y)^2}{1 + (1 - y)^2} a_3 \Big] \,  ,
\end{equation}
where $a_1$ and $a_3$ are the ratios of structure functions, which can be written in terms of parton distribution functions (PDFs),
 \begin{eqnarray}
 \label{eq:a1-a3-SM}
 a_1 &=& \frac{2 \sum_q e_q C_{1q} (q + \bar{q})}{\sum_q e_q^2 (q + \bar{q})} ,\nonumber\\
 a_3 &=& \frac{2 \sum_q e_q C_{2q} (q - \bar{q})}{\sum_q e_q^2 (q + \bar{q})} \, .
 \end{eqnarray}
Here $C_{1q} = 2 g^e_A g^q_V$ and $C_{2q} = 2 g^e_V g^q_A$ are the axial-vector vector (AV) and vector axial-vector (VA) combinations of the electron and the quark weak couplings, respectively.
The SM couplings are
\begin{eqnarray}
\{ g^e_V, g^u_V, g^d_V\} &=& \{ - \frac{1}{2} + 2 \sin^2\theta_W, \frac{1}{2} - \frac{4}{3}\sin^2\theta_W \, , \nonumber\\
&& \ - \frac{1}{2} + \frac{2}{3}\sin^2\theta_W\} , \nonumber\\
\{ g^e_A, g^u_A, g^d_A\} &=& \{ - \frac{1}{2}, \frac{1}{2}, -\frac{1}{2} \} \, ,
\end{eqnarray}
 where $\theta_W$ is the Weinberg angle. 
 
{\em Sensitivity of PVES asymmetry to the dark photon:} The motivation for the existence of a dark photon in the general context of the search for dark matter was recently reviewed by Filippi and Napoli~\cite{Filippi:2020kii}. The idea began with a 
proposal~\cite{Holdom:1985ag,Fayet:1980ad} of a spin-one gauge boson mixing kinematically with the $U(1)_Y$ boson in the Standard Model. It was proposed that it might provide a portal to other hidden particles through this mixing. 
\begin{equation}
\mathcal{L}  \supset - \frac{1}{4} F'_{\mu\nu} F'^{\mu\nu} + \frac{m^2_{A'}}{2} A'_{\mu} A'^{\mu} + \frac{\epsilon}{2 \cos\theta_W} F'_{\mu\nu} B^{\mu\nu} \, .
\end{equation}
We use $A'$ and $\bar{Z}$ to denote the unmixed versions of the dark photon and the SM neutral weak boson, respectively.

After diagonalizing the mixing term through field redefinitions, the physical masses of the Z boson and the dark photon are~\cite{Kribs:2020vyk}
\begin{eqnarray}
M^2_{Z, A_D} &=& \frac{m_{\bar{Z}}^2}{2} [ 1 + \epsilon_W^2 + \rho^2 \nonumber\\
&& \pm {\rm sign}(1-\rho^2) \sqrt{(1 + \epsilon_W^2 + \rho^2)^2 - 4 \rho^2} ] \, ,
\end{eqnarray}
where
\begin{eqnarray}
\epsilon_W &=& \frac{\epsilon \tan \theta_W}{\sqrt{1 - \epsilon^2/\cos^2\theta_W}} ,\nonumber\\
\rho &=& \frac{m_{A'}/m_{\bar{Z}}}{\sqrt{1 - \epsilon^2/\cos^2\theta_W}} \, .
\end{eqnarray}
The SM couplings of the $Z$-boson, $C_{\bar{Z}}^v = \{ g^e_V, g^u_V, g^d_V\}$ and $C_{\bar{Z}}^a = \{ g^e_A, g^u_A, g^d_A\}$, will be modified because of the kinetic mixing~\cite{Kribs:2020vyk}
\begin{eqnarray}
\label{eq:C_Z}
C_Z^v &=& (\cos\alpha - \epsilon_W \sin\alpha) C_{\bar{Z}}^v + 2 \epsilon_W \sin\alpha \cos^2 \theta_W C_{\gamma}^v ,\nonumber\\
C_Z^a &=& (\cos\alpha - \epsilon_W \sin\alpha) C_{\bar{Z}}^a ,
\end{eqnarray}
where 
$C_{\gamma}^v = \{ C^e_{\gamma}, C^u_{\gamma}, C^d_{\gamma}\} = \{ -1, 2/3, - 1/3 \}$. 
The couplings of the physical dark photon $A_D$ to SM particles are given by
\begin{eqnarray}
\label{eq:C_AD}
C_{A_D}^v &=& - (\sin\alpha + \epsilon_W \cos\alpha) C_{\bar{Z}}^v + 2 \epsilon_W \cos\alpha \cos^2 \theta_W C_{\gamma}^v ,\nonumber\\
C_{A_D}^a &=& - (\sin\alpha + \epsilon_W \cos\alpha) C_{\bar{Z}}^a 
\, .
\end{eqnarray}
Here $\alpha$ is the $\bar{Z}-A'$ mixing angle,
\begin{eqnarray}
\tan \alpha &=& \frac{1}{2\epsilon_W} \Big[ 1 - \epsilon^2_W - \rho^2 \nonumber\\
&& - {\rm sign}(1-\rho^2) \sqrt{4\epsilon_W^2 + (1 - \epsilon_W^2 - \rho^2)^2} \Big] \, . 
\end{eqnarray}

The strongest constraints on $\epsilon$ come from the  NA64~\cite{Banerjee:2019pds} and {\em BABAR} experiments~\cite{BaBar:2017tiz}, leading to $\epsilon\le 10^{-3}$ for $M_{A_D} \le 8\ {\rm GeV}$. This limit could possibly be weakened if the detailed structure of the dark sector were taken into account~\cite{Essig:2009nc}.
The current limit in connection with electroweak precision observables (EWPO)~\cite{Hook:2010tw, Curtin:2014cca} leads to $\epsilon \le 0.03$ for dark photon mass up to $M_Z$, while the upcoming high-luminosity LHC run (HL-LHC) is expected to place a very strong constraint of $\epsilon \le 10^{-6}$ in this mass region~\cite{Curtin:2014cca}. 
The exclusion limits from recent $e^{-}p$ DIS analyses are either compatible with~\cite{Thomas:2021lub} or slightly stronger than the EWPO bound, $\epsilon \le 0.02$ for $M_{A_D} < 10\ {\rm GeV}$~\cite{Kribs:2020vyk, Yan:2022npz}, which become weaker as the dark photon mass increases. Especially in the heavy mass region, to which we restrict our attention, the upper limit on $\epsilon$ from the DIS determination will go above 0.1 when $M_{A_D} > M_Z$~\cite{Kribs:2020vyk}.

However, in the present context it is crucial that the dark photon will also contribute to the PVES asymmetry in DIS because of its axial-vector couplings to the electron and the quarks.
The double differential cross section can be expressed as
\begin{eqnarray}
\frac{d^2 \sigma}{dx dy} 
&=& \frac{4\pi \alpha^2 s}{Q^4}
\Big(
[x y^2 F_1^{\gamma} + f_1(x,y) F_2^{\gamma}]  \nonumber\\
&&
 - \frac{1}{\sin^2 2\theta_W}\frac{Q^2}{Q^2 + M_Z^2} (C_{Z,e}^v - \lambda C_{Z,e}^a) \times \nonumber\\
&& [x y^2 F_1^{\gamma Z} + f_1(x,y) F_2^{\gamma Z} - \lambda x y (1-\frac{y}{2}) F_3^{\gamma Z}]  \nonumber\\
&&
-  \frac{1}{\sin^2 2\theta_W} \frac{Q^2}{Q^2 + M_{A_D}^2} (C_{A_D,e}^v - \lambda C_{A_D,e}^a) \times \nonumber\\
&& [x y^2 F_1^{\gamma A_D} + f_1(x,y) F_2^{\gamma A_D} - \lambda x y (1-\frac{y}{2}) F_3^{\gamma A_D}]
\Big) \, , \nonumber\\
\end{eqnarray}
where  $f_1(x,y) = 1 - y - xyM/2E$ and $\lambda = + 1 (-1)$ represents positive (negative) initial electron helicity. 
 For positron scattering, the cross sections can be obtained with $C_{Z,e}^a$ and $C_{A_D,e}^a$ being replaced by $-C_{Z,e}^a$ and $-C_{A_D,e}^a$, respectively~\cite{Anselmino:1993tc}.

Since the purely electromagnetic cross section does not contribute to the asymmetry, 
the numerator receives contributions from $\gamma-Z$ and $\gamma-A_D$ interference terms,
\begin{eqnarray}
\label{eq:APV-Z-AD}
A_{\rm PV} &=& 
 \frac{Q^2}{2 \sin^2 2\theta_W (Q^2 + M^2_Z)} 
  \Big[ a_1^{\gamma Z} + \frac{1 - (1 - y)^2}{1 + (1 - y)^2} a_3^{\gamma Z} \nonumber\\
&& + \frac{Q^2 + M_Z^2}{Q^2 + M_{A_D}^2} (a_1^{\gamma A_D} + \frac{1 - (1 - y)^2}{1 + (1 - y)^2} a_3^{\gamma A_D} )\Big] ,\nonumber\\
\end{eqnarray}
where $a^{\gamma Z}_1 (a^{\gamma A_D}_1)$ and $a^{\gamma Z}_3 (a^{\gamma A_D}_3)$ have the same form as Eq.~(\ref{eq:a1-a3-SM}),
with the corresponding $C^Z_{1q} (C^{A_D}_{1q})$ and $C^Z_{2q} (C^{A_D}_{2q})$ defined by the physical couplings given in Eqs.~(\ref{eq:C_Z}-\ref{eq:C_AD}).
For $Q^2 \ll M_Z^2$, $A_{\rm PV}$ can  be rewritten in terms of the Fermi constant $G_F$ using the relation
\begin{equation}
\label{eq:G_F}
\frac{Q^2}{2 \sin^2 2 \theta_W(Q^2 + M_Z^2)} = \frac{G_F Q^2}{4\sqrt{2} \pi \alpha} \, .
\end{equation}
From Eq.~(\ref{eq:APV-Z-AD}) the effect of both $Z$ and $A_D$ exchange is given by the effective couplings 
\begin{eqnarray}
\label{eq:C1q-C2q}
C_{1q} &=& C^Z_{1q} + \frac{Q^2 + M_Z^2}{Q^2 + M_{A_D}^2} C^{A_D}_{1q} = C^{\rm SM}_{1q} ( 1 + R_{1q} ),\nonumber\\
C_{2q} &=& C^Z_{2q} + \frac{Q^2 + M_Z^2}{Q^2 + M_{A_D}^2} C^{A_D}_{2q} = C^{\rm SM}_{2q} ( 1 + R_{2q} ) \,  ,
\end{eqnarray}
with $R_{1q}$ and $R_{2q}$ characterizing the corrections to the SM couplings, arising from the effects of a dark photon.

For this study, the dark photon parameter space corresponding to $\epsilon \le 0.2$ in the $(\epsilon, M_{A_D})$ plane is of most interest, because it has not been fully excluded by the existing constraints.
The parameters in the ``eigenmass repulsion" region, very near the $Z$-boson mass, are not accessible~\cite{Kribs:2020vyk}.

{\em PREX:}  As explained earlier, the PREX experiment returned a value for the difference of proton and neutron radii in $^{208}$Pb that was considerably larger than expected from standard nuclear structure calculations. This may be interpreted as implying that the slope of the symmetry energy as a function of density is considerably larger than hitherto believed. It has been suggested that this would have important implications for the structure of nuclei away from stability~\cite{Pineda:2021shy}, as well as the properties of neutron stars -- notably their surface thickness and radii~\cite{Reed:2021nqk,Biswal:2021mlf}.

\begin{figure}[!h]
\includegraphics[width=\columnwidth]{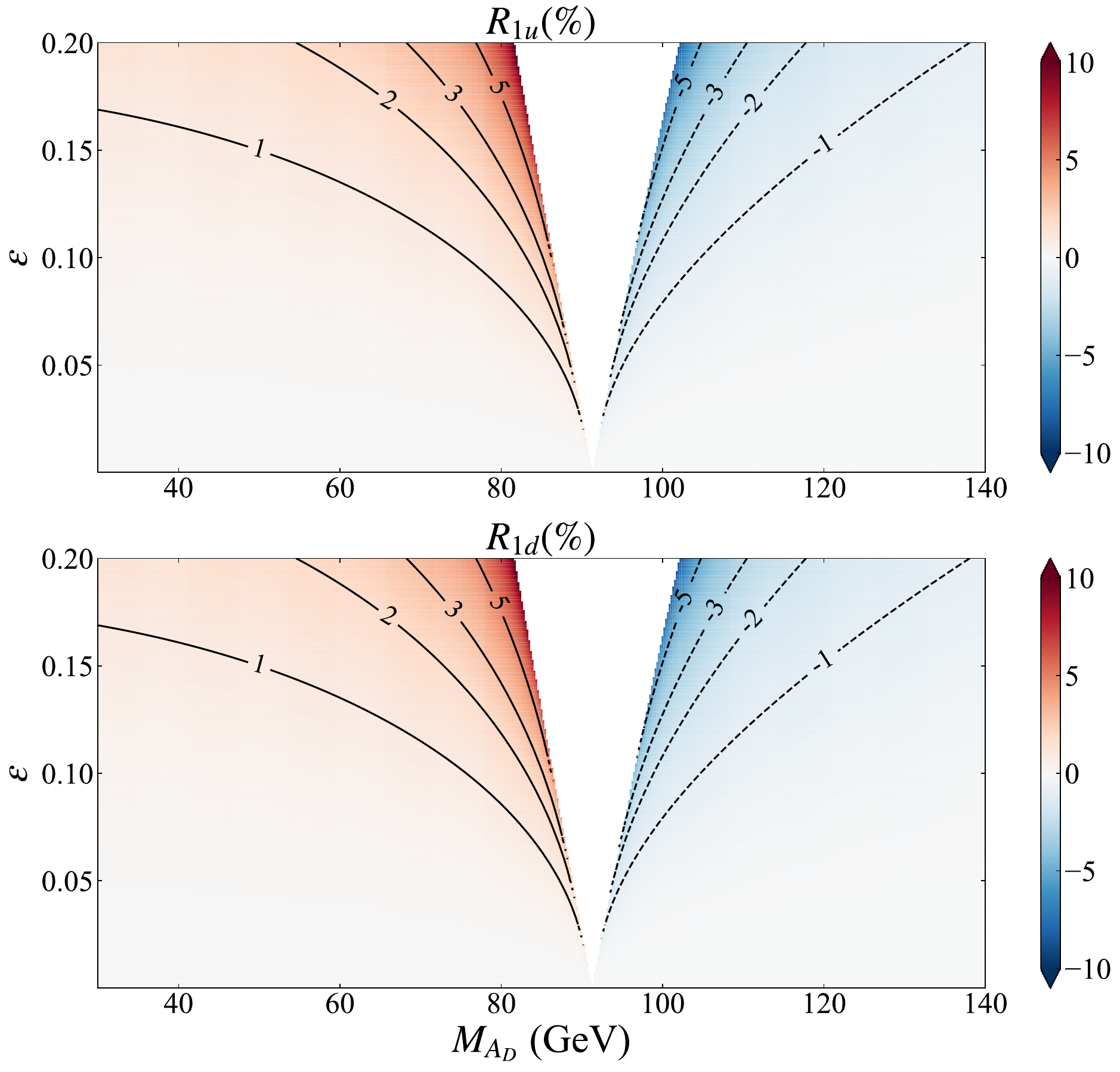}
\vspace*{-0.2cm}
\caption{The correction factors $R_{1u}$ and $R_{1d}$ at $Q^2 = 0.00616\ {\rm GeV}^2$, appropriate to the PREX-II experiment.
The gap on the $\epsilon-M$ plane is not accessible because of ``eigenmass repulsion" associated with the $Z$ mass.}
\label{fig:R1u-R1d-Q2low}
\end{figure}

We first consider the case of very low momentum transfer, $Q^2 = 0.00616\ {\rm GeV}^2$, which is relevant for elastic scattering in the PREX experiment~\cite{PREX:2021umo}.
In Fig.~\ref{fig:R1u-R1d-Q2low} we show the correction $R_{1q}$ in the $(\epsilon, M_{A_D})$ plane.
At this low scale, the correction to $C_{1q}$ can be as large as several percent when the dark photon parameters approach the ``eigenmass repulsion" region. We observe that a correction of this size is very significant in the context of the PREX measurement. For example, a value of $R_{1q}$ of order 4\% would lead to a decrease in the deduced neutron radius of Pb, which would entirely eliminate any tension with the theoretically prefered values. Finally on this topic, we observe that at low momentum transfer the dark photon gives rise to changes in the up and down quark couplings that are roughly independent of flavor and so cannot simply be represented by a change in the Weinberg angle.
%
%
%

\begin{figure}[!h]
\includegraphics[width=\columnwidth]{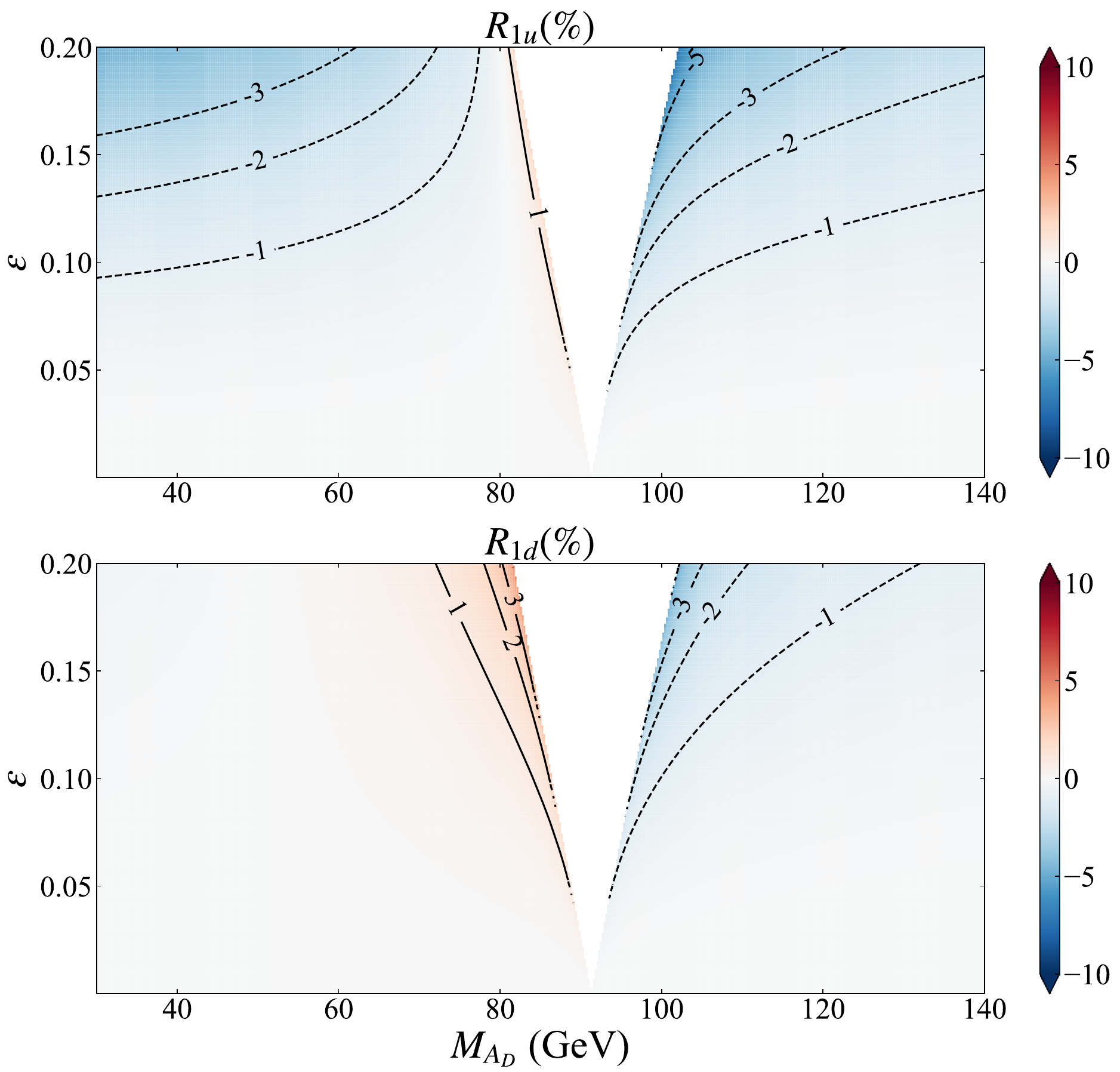}
\vspace*{-0.2cm}
\caption{The correction factors $R_{1u}$ and $R_{1d}$ at $Q^2 = M^2_Z$.}
\label{fig:R1u-R1d-Q2high}
\end{figure}

\begin{figure}[!h]
\includegraphics[width=\columnwidth]{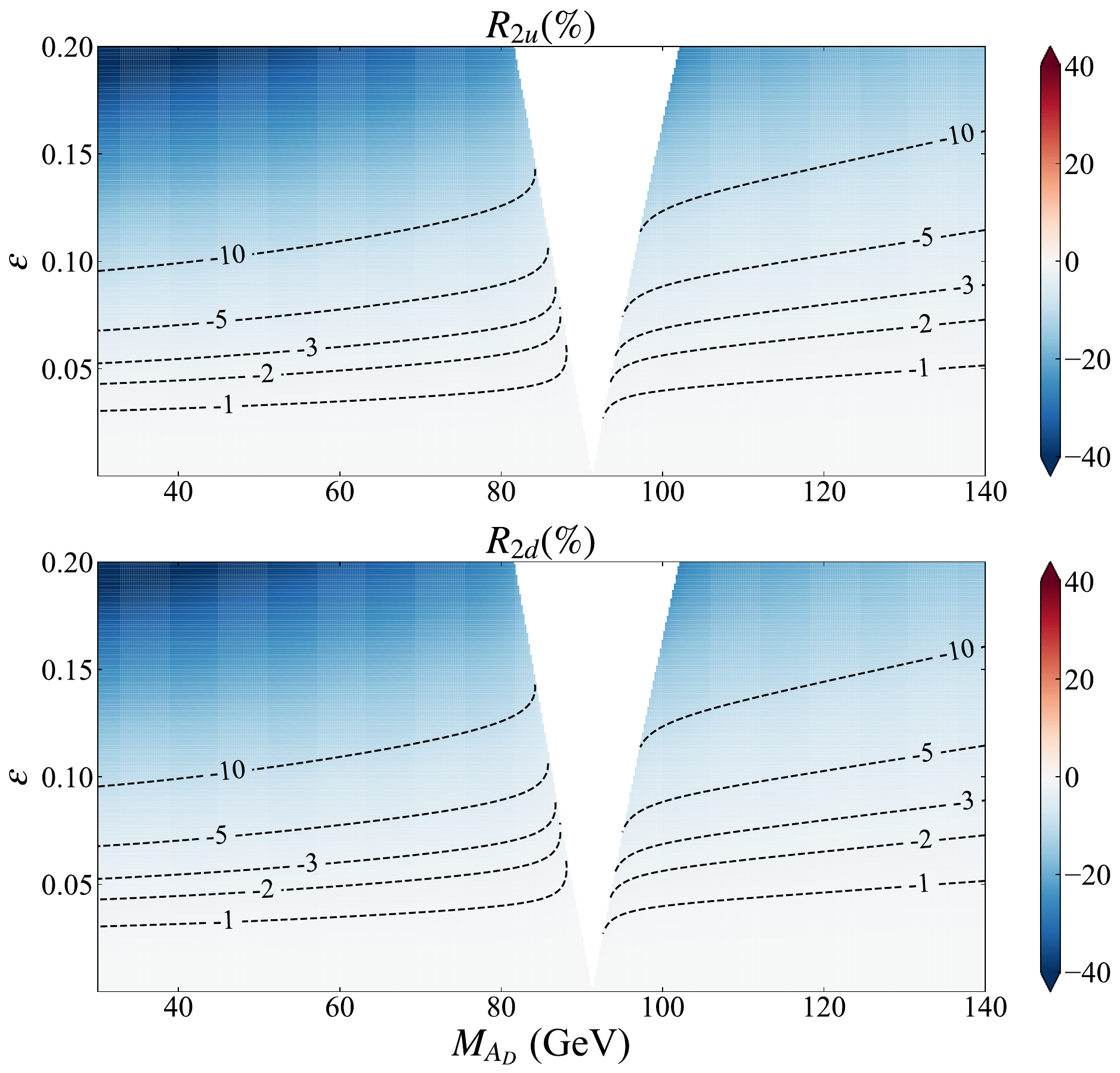}
\vspace*{-0.2cm}
\caption{The correction factors $R_{2u}$ and $R_{2d}$ at $Q^2 = M^2_Z$.}
\label{fig:R2u-R2d-Q2high}
\end{figure}

{\em Deep inelastic scattering at high $Q^2$:}  Next we investigate the dark photon effects at the much higher momentum scales associated with DIS. This is especially relevant for the measurements taken at HERA, where the large values of $Q^2$ accessible in the collider made the measurement of the C-odd valence quark distributions possible. Both $C_{1q}$ and $C_{2q}$ make significant contributions to the PVDIS asymmetry in this region.
The correction factors $R_{1q}$ and $R_{2q}$ at $Q^2 = M^2_Z$ are shown 
in Figs.~\ref{fig:R1u-R1d-Q2high} and \ref{fig:R2u-R2d-Q2high}, respectively.
At this scale the behaviour of the corrections to $C_{1u}$ and $C_{1d}$ are qualitatively very different. While the corrections $C_{1q}$ are relatively small in this case, those for $C_{2q}$ tend to be negative over the entire region and can be large in magnitude even if the kinetic mixing parameter $\epsilon$ is relatively small.

As values of $\epsilon$ as large as 10-15\% are not excluded in the dark photon mass region above 70 GeV, the results shown in Figs.~\ref{fig:R1u-R1d-Q2high} and \ref{fig:R2u-R2d-Q2high} suggest that the errors in the extraction of valence parton distribution functions from high-$Q^2$ data at HERA could be as large as 10\% or more, were a dark photon to exist. 

In the future, the kinematic coverage of the electron-ion collider (EIC)~\cite{NAS} planned in the United States, will allow considerable improvement in our knowledge of the PDFs over a wide range of $x$ and $Q^2$, as explained in the EIC Yellow Report\cite{AbdulKhalek:2021gbh}. However, without access to a positron beam, which is not yet certain, the direct determination of $F_3$ will not be possible. On the other hand, with improved the measurements of the sea at the EIC and very accurate measurements of $F_{1,2}$ at both the EIC and JLab 12 GeV, the accuracy with which the valence PDFs are known may be expected to improve significantly.

{\em $g^{eq}_{AA}$:}  One of the key experiments planned with the SoLID detector at 
JLab~\cite{Accardi:2020swt} is the first measurement of the parameter $g^{eq}_{AA}$~\cite{Erler:2013xha}. 
At leading order this is the Standard Model coupling of an electron to a quark with an axial current at each vertex, otherwise known as $C_{3q}$. 
With the effect of the dark photon, $C_{3q}$ becomes
\begin{equation}
C_{3q} = C^Z_{3q} + \frac{Q^2 + M_Z^2}{Q^2 + M_{A_D}^2} C^{A_D}_{3q} = C^{\rm SM}_{3q} ( 1 + R_{3q} ) \, .
\end{equation}
Like the coefficients $C_{1,2}$ discussed above, these are fundamental coefficients in the SM and any confirmed deviation would signal new physics. While it has been tested at CERN for muons~\cite{Argento:1982tq}, at an accuracy of order 25\%, the axial-axial coupling has never been measured for electrons. 
The ideal experiment to determine this coupling is to measure the difference in unpolarized electron and positron scattering on the deuteron~\cite{Zheng:2021hcf}
\begin{equation}
A^{e^+ e^-}_d = - \frac{3 G_F Q^2 Y}{2\sqrt{2}\pi\alpha}
\frac{R_V(2 g^{eu}_{AA} - g^{ed}_{AA})}{5+4 R_C + R_S} \, ,
\end{equation}
where following Ref.~\cite{Zheng:2021hcf} $g^{eq}_{AA}$ is defined to incorporate higher order radiative corrections, including, for example, two-photon exchange.

\begin{figure}[!h]
\includegraphics[width=\columnwidth]{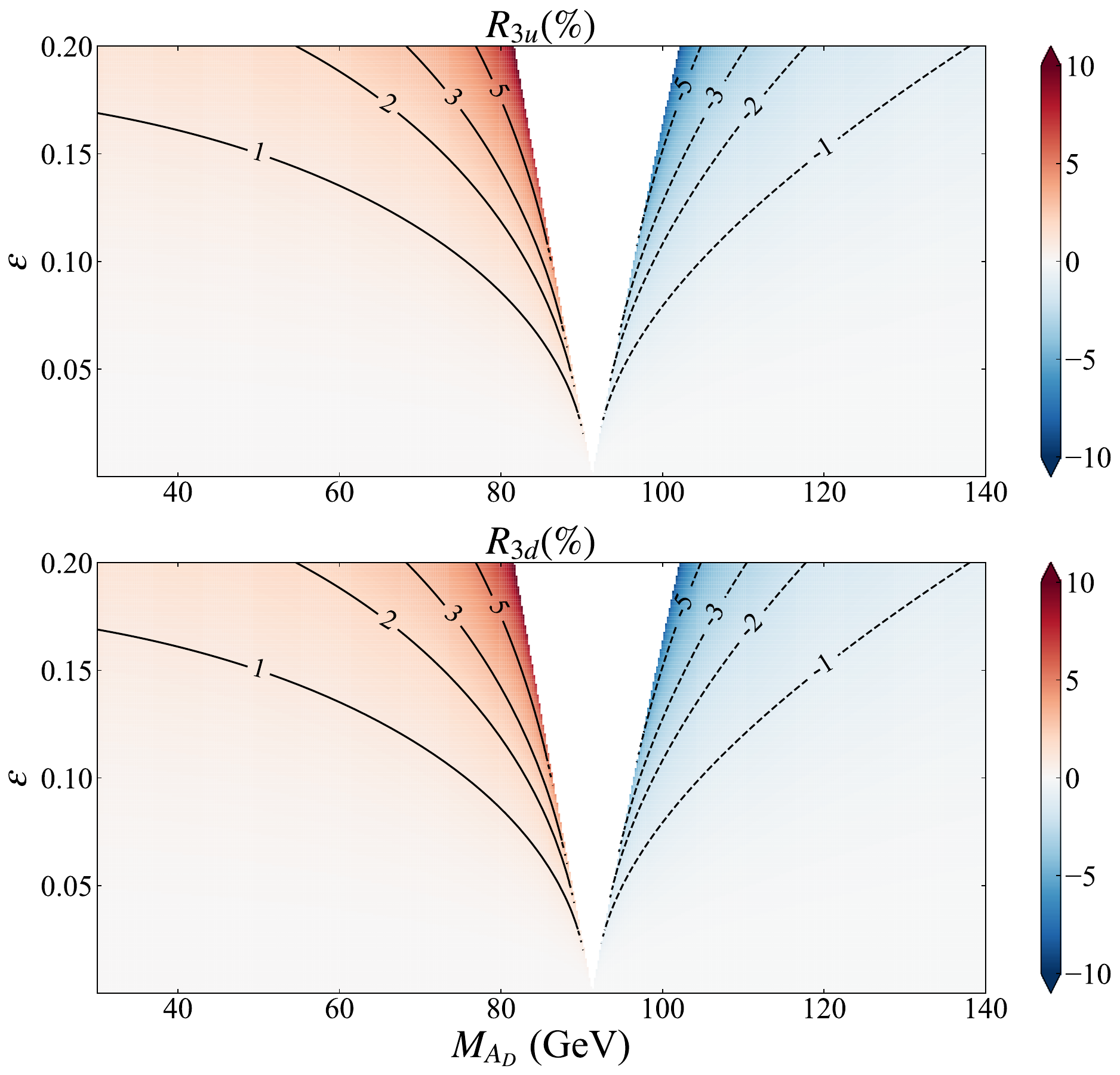}
\vspace*{-0.2cm}
\caption{The correction factors $R_{3u}$ and $R_{3d}$ at $Q^2 = 10\ {\rm GeV}^2$.}
\label{fig:R3u-R3d-Q2-10GeV2}
\end{figure}

As shown in Fig.~\ref{fig:R3u-R3d-Q2-10GeV2}, our calculations suggest that there are kinematic regions where the dark photon could lead to deviations as large as $5\%$ from SM expectations at the scale $Q^2 = 10\ {\rm GeV}^2$, appropriate to possible experiments at JLab.

 {\em Summary:} \, We have calculated the dark photon contributions to parity-violating electron scattering (PVES).
These contributions  are characterized  by the corrections to the standard model couplings 
$C_{1q}, \, C_{2q}$ and $C_{3q}$. For elastic scattering we showed that there could be a relatively large correction to the neutron radius of the Pb nucleus deduced from the PVES measurement of PREX. On the other hand, the allowed changes are sufficiently small that they have no effect on the interpretation of the Qweak experiment. In DIS at very high $Q^2$, of relevance to HERA, the dark photon could induce substantial corrections to the valence parton distribution functions deduced from the DIS data. Finally, the electron-positron asymmetry in DIS offers direct access to the combination $2C_{3u}-C_{3d}$, where effects as large as $5\%$ are possible.

These results suggest that it would be extremely valuable to have a dedicated program to test for the existence of a dark photon.

\vspace{0.3cm}
We would like to acknowledge helpful correspondence with Xiaochao Zheng. This work was supported by the University of Adelaide and the Australian Research Council through the Centre of Excellence for Dark Matter Particle Physics (CE200100008) and Discovery Project DP180100497 (AWT).


\end{document}